\documentclass[sigconf]{acmart}
\makeatletter
\def\@ACM@checkaffil{%
    \if@ACM@instpresent\else
    \ClassWarningNoLine{\@classname}{No institution present for an affiliation}%
    \fi
    \if@ACM@citypresent\else
    \ClassWarningNoLine{\@classname}{No city present for an affiliation}%
    \fi
    \if@ACM@countrypresent\else
        \ClassWarningNoLine{\@classname}{No country present for an affiliation}%
    \fi
}
\makeatother

\AtBeginDocument{%
  \providecommand\BibTeX{{%
    \normalfont B\kern-0.5em{\scshape i\kern-0.25em b}\kern-0.8em\TeX}}}

\copyrightyear{2024}
\acmYear{2024}
\setcopyright{rightsretained}
\acmConference[ASE '24]{39th IEEE/ACM International Conference on Automated Software Engineering }{October 27-November 1, 2024}{Sacramento, CA, USA}
\acmBooktitle{39th IEEE/ACM International Conference on Automated Software Engineering (ASE '24), October 27-November 1, 2024, Sacramento, CA, USA}
\acmDOI{10.1145/3691620.3695033}
\acmISBN{979-8-4007-1248-7/24/10}

\usepackage{url}
\usepackage{pifont} 
\usepackage{booktabs}
\usepackage{multirow}
\usepackage{tabularx}
\usepackage{xcolor}
\usepackage{xspace}
\usepackage{listings}
\usepackage{xstring}
\usepackage[htt]{hyphenat}
\usepackage{subcaption}
\usepackage{makecell}
\usepackage{colortbl}
\usepackage{paralist}
\usepackage{soul}

\begin{document}

\title[Identifying Systematic Problems with Semantic Data Slicing]{What Is Wrong with My Model?\\
Identifying Systematic Problems with Semantic Data Slicing}

\author{Chenyang Yang}
\affiliation{%
  \institution{Carnegie Mellon University}
}

\author{Yining Hong}
\affiliation{%
  \institution{Carnegie Mellon University}
}

\author{Grace A. Lewis}
\affiliation{%
  \institution{Carnegie Mellon Software Engineering Institute}
}

\author{Tongshuang Wu}
\affiliation{%
  \institution{Carnegie Mellon University}
}

\author{Christian K{\"a}stner}
\affiliation{%
  \institution{Carnegie Mellon University}
}

\renewcommand{\shortauthors}{Yang, et al.}

\begin{abstract}
Machine learning models make mistakes, yet sometimes it is difficult to identify the systematic problems behind the mistakes.
Practitioners engage in various activities, including error analysis, testing, auditing, and red-teaming,
to form hypotheses of what can go (or has gone) wrong with their models.
To validate these hypotheses, practitioners employ data slicing to identify relevant examples.
However, traditional data slicing is limited by available features and programmatic slicing functions.
In this work, we propose SemSlicer, a framework that supports semantic data slicing, which identifies a semantically coherent slice, without the need for existing features.
SemSlicer uses Large Language Models to annotate datasets and generate slices from any user-defined slicing criteria.
We show that SemSlicer generates accurate slices with low cost, allows flexible trade-offs between different design dimensions,
reliably identifies under-performing data slices, and helps practitioners identify useful data slices that reflect systematic problems.
\end{abstract}

\def\BibTeX{{\rm B\kern-.05em{\sc i\kern-.025em b}\kern-.08em
    T\kern-.1667em\lower.7ex\hbox{E}\kern-.125emX}}

\maketitle

\newcommand{\toolname}[0]{\textsc{SemSlicer}}
\newcommand\circleone{\ding{192}}
\newcommand\circletwo{\ding{193}}
\newcommand\circlethree{\ding{194}}
\newcommand\circlefour{\ding{195}}
\newcommand\circlefive{\ding{196}}
\newcommand\circlesix{\ding{197}}

\newcommand{\DGone}{{\texttt{accuracy}}}
\newcommand{\DGtwo}{{\texttt{human-effort}}}
\newcommand{\DGthree}{{\texttt{compute}}}
\newcommand{\DGfour}{{\texttt{latency}}}

\newcommand{\nbc}[3]{
 {\colorbox{#3}{\bfseries\sffamily\scriptsize\textcolor{white}{#1}}}
 {\textcolor{#3}{\sf\small$\blacktriangleright$\textit{#2}$\blacktriangleleft$}}
 }

 \newcommand\todo[1]{\nbc{TODO}{#1}{red}}
\newcommand{\cyang}[1]{\nbc{CY}{#1}{teal}}
\newcommand{\sherry}[1]{\nbc{SW}{#1}{purple}}
\newcommand{\grace}[1]{\nbc{GL}{#1}{blue}}

\definecolor{mygreen}{rgb}{0,0.6,0}
\definecolor{mygray}{rgb}{0.5,0.5,0.5}
\definecolor{mymauve}{rgb}{0.58,0,0.82}

\definecolor{rose}{rgb}{1, 0.89, 0.88}
\definecolor{lightgreen}{HTML}{c7e9c0}
\definecolor{darkgreen}{HTML}{a1d99b}

\newcommand{\markpoor}{\cellcolor{rose}} %
\newcommand{\markgood}{\cellcolor{lightgreen}}
\newcommand{\markbest}{\cellcolor{darkgreen}}

\section{Introduction}
\begin{figure}
    \centering
    \includegraphics[width=0.9\linewidth]{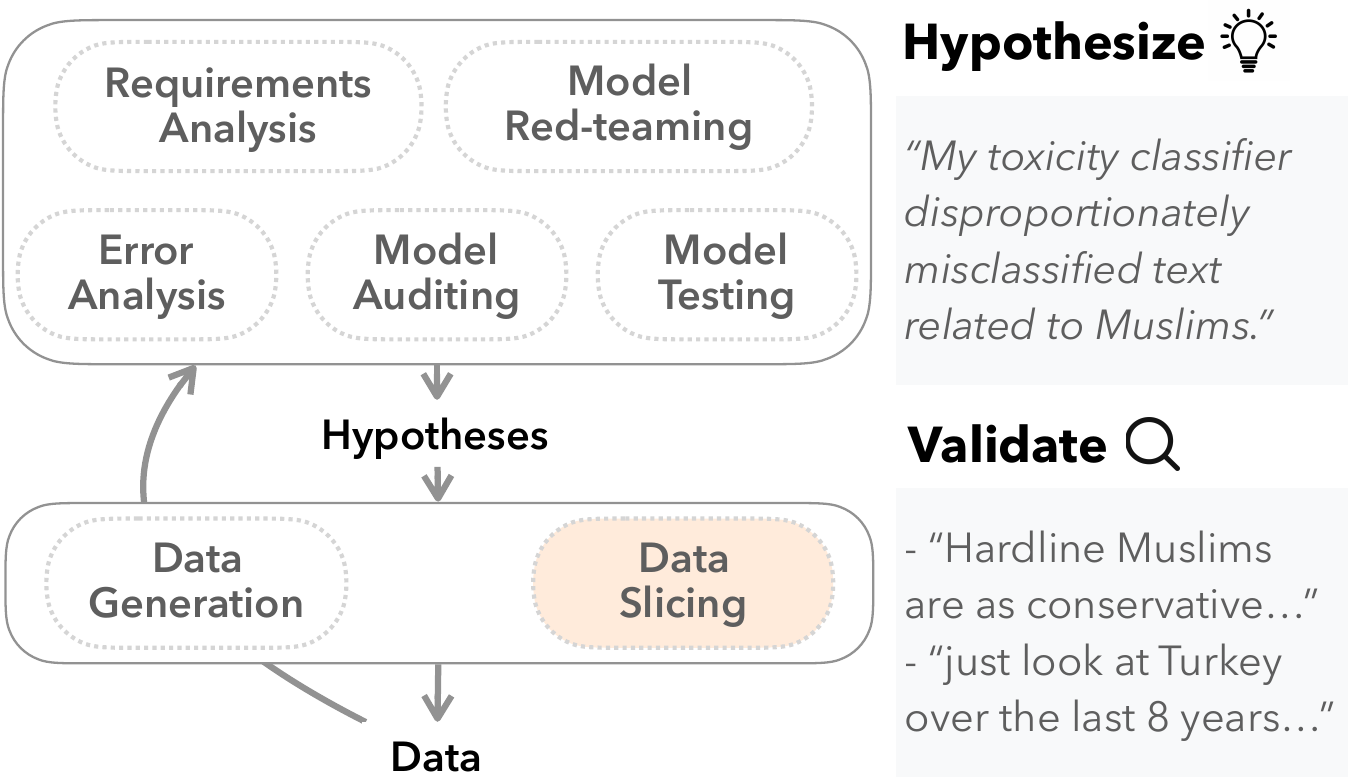}
    \caption{
    \small
    ML model quality assurance involves two stages: (1) Hypothesize and (2) Validate. 
    Many activities focus on creating hypotheses, either explicitly (requirements analysis, error analysis) or implicitly in the process (testing, auditing, red-teaming).
    \textbf{Data slicing} helps validate the produced hypotheses by identifying \textit{additional} relevant examples, often from evaluation and production data.
    }
    \label{fig:eval_flow}
    \vspace{-10pt}
\end{figure}

Machine learning models exhibit undesired behaviors, yet it is often hard to identify systematic problems behind individual errors.
Models have been found to perform worse on under-represented subgroups~\cite{pmlr-v139-WILDS}, exhibit unintended biases~\cite{sap-etal-2019-risk}, and produce harmful content~\cite{weidinger2021ethical},
which can cause project failures, media controversies, and even lawsuits once they are integrated into software products~\cite{consultantsnews}.
Academia and industry have spent significant effort to help identify these problems, through activities such as error analysis, testing, auditing, and red-teaming~\cite{Naik2018StressTE, ribeiro-lundberg-2022-adaptive, metaxa21auditing, feffer2024red}.
All these activities (Figure~\ref{fig:eval_flow}, top) curate \textit{individual errors}, %
but more importantly, also create concrete \textit{hypotheses} about the underlying systematic problem.\footnote{Alternatively, the hypotheses can also be formed top-down through explicit \textbf{requirements analysis}~\cite[e.g.,][]{improvingdomainspecs, weaver}, which is less common in current ML practice.}

As our running example, suppose an ML practitioner is conducting error analysis on their toxicity classifier~\cite{hosseini2017deceiving}.
The practitioner observed that in a few examples, the model classifies non-toxic text mentioning Muslims as toxic.
From these individual errors, they hypothesize that the systematic problem behind the errors might be that \textit{``the classifier disproportionately misclassified text related to Muslims.''}
To understand whether their hypotheses hold, the practitioner needs to \textit{validate} the hypotheses on more data points (Figure~\ref{fig:eval_flow}, bottom).

To validate their hypotheses, developers need to (a) conduct synthetic data generation~\cite{he-etal-2023-targeted} to create more data points or (b) identify relevant data points from existing data.
\textit{Data slicing}~\cite{slicefinder, errudite}, an example of the latter type of technique, identifies a subset of examples sharing common characteristics from existing data.
Data slicing often assumes access to existing relevant features, which might not always be available for users' slicing criteria of interest. For example, it is unlikely that the input data is readily labeled with whether they are related to Muslims.

In practice, developers try to create additional features to augment datasets for data slicing but are limited by \textit{how} they can create such features:
Existing practices mostly apply \textit{programmatic} slicing (see statistics in Table~\ref{tab:zeno}), often implemented as simple Python programs, to identify slices.
For our example hypothesis (\textit{``the classifier disproportionately misclassified text related to Muslim''}), a simple implementation is to use a regular expression to search for the pattern ``muslim|islam'' (Figure~\ref{fig:slicing_func}, top).
However, this simple implementation leaves out many related but more nuanced examples (e.g., when the examples mention ``Saudi Arabia'', ``Turkey'', as shown in Figure~\ref{fig:slicing_func}, bottom).
Writing a regex-based slicing function to cover all these different cases can be a laborious process, and it is hard to enumerate these patterns to begin with.
As we will show later in Section~\ref{sec:background}, this is a common theme in data slicing---developers often want to slice based on criteria that are hard to implement with programs.
Limitations of existing \textit{programmatic} slicing methods fundamentally constrain the applicability of data slicing.

In this work, we propose \toolname{}, a framework that supports semantic data slicing using Large Language Models (LLMs).
\textit{Semantic data slicing} identifies a {semantically coherent} subset of examples, without the need for existing features as in programmatic data slicing.
The key insight of \toolname{} is that with appropriate prompts, LLMs can act as slicing functions for any user-specified slicing criteria. 
These slicing functions formed by an LLM and a prompt are able to cover semantically coherent examples of different surface patterns. %
To assist users in creating slicing prompts, we design a highly configurable prompt construction pipeline in \toolname{}, which allows different levels of machine and human intervention.

To use \toolname{}, a user needs to specify (a) a slicing criterion, expressed as a keyword or phrase (e.g., ``Muslim''), (b) a dataset to slice (e.g., a test dataset or recent production data), and (c) the exact configuration for slicing (i.e., choosing among various forms of auto-optimization to trade off cost and accuracy, as we will explain in Section~\ref{sec:approach}). 
\toolname{} will then (1) generate and automatically optimize a slicing prompt (e.g., generating few-shot examples), and (2) annotate the entire dataset and return a relevant slice.
Applying \toolname{} to our running example (Figure~\ref{fig:slicing_func}, bottom), the slicing criterion can be as simple as ``Muslim'' (line~\ref{code:slicing_criteria}), from which \toolname{} will produce a slicing prompt (line~\ref{code:slicing_prompt}) that can be used to obtain the corresponding slice (line~\ref{code:semslicer_result}).

Besides model debugging as in our example, semantic data slicing is widely applicable to many activities in ML engineering, such as assisted data curation, systematic model evaluation against requirements, and model monitoring (as we will discuss in Section~\ref{sec:background}).
To accommodate different use cases, \toolname{} provides a rich set of configuration options, such that users can make flexible trade-offs:  
adjust the models used in different components according to compute budgets; change whether and how to produce few-shot examples to trade off between compute resources and accuracy; and collaborate with LLMs in producing instructions, to provide additional human oversight. 
Our evaluation shows that \toolname{} can produce accurate slicing functions for a wide range of slicing criteria, allows flexible trade-offs between cost and accuracy, and is useful for model evaluation (Section~\ref{sec:intrinsic_eval}).

In summary, our work makes the following contributions:
\begin{compactitem}
    \item A comprehensive view of the landscape of data slicing in ML engineering.
    \item A highly configurable framework, \toolname{}, that supports semantic data slicing for diverse use cases.\footnote{\toolname{} is available open-source at \url{https://github.com/malusamayo/SemSlicer}.}
    \item An extensive evaluation of \toolname{} that shows it can generate \textit{accurate} slicing functions,  allows \textit{flexible} trade-offs,
    and is \textit{useful} for model evaluation.
\end{compactitem}

\let\origthelstnumber\thelstnumber
\makeatletter
\newcommand*\Suppressnumber{%
  \lst@AddToHook{OnNewLine}{%
    \let\thelstnumber\relax%
     \advance\c@lstnumber-\@ne\relax%
    }%
}

\newcommand*\Reactivatenumber{%
  \lst@AddToHook{OnNewLine}{%
   \let\thelstnumber\origthelstnumber%
   \advance\c@lstnumber\@ne\relax}%
}

\begin{figure}[t]
\lstset{
        extendedchars=true,
        basicstyle=\footnotesize\ttfamily,
        showstringspaces=false,
        showspaces=false,
        numbers=left,
        numberstyle=\scriptsize\ttfamily,
        numbersep=7pt,
        xleftmargin=1em,
        framexleftmargin=2em,
        tabsize=2,
        breaklines=true,
        showtabs=false,
        captionpos=b,
        language=python, 
        frame=lines, 
        commentstyle=\color{mygreen},
        escapeinside={\%}{\%},          
        keywordstyle=\color{blue},       
        stringstyle=\color{mymauve},     
        literate={\$}{{\textcolor{mymauve}{\$}}}1
}
\begin{lstlisting}[language=Python]
# Programmatic data slicing detects texts with simple patterns but generalizes poorly.
import re
data = load_training_data()
%\label{code:slicing_regex}%pattern = r"muslim|islam"

%\label{code:programmatic_slicing}%def regex_slicing(x: str) -> bool:
    return bool(re.search(pattern, x, re.IGNORECASE))

m_slice = data[data['text'].map(regex_slicing)]
m_slice['text'].sample(2)%\Suppressnumber%
%$\mid$% '''there's the muslim lady in jail for trying to kill 
%$\mid$%      people with a golf club in CTC
%$\mid$%    Europe is in the process of submitting to Islam a 
%$\mid$%      bit more every year.'''%\Reactivatenumber%
\end{lstlisting}
\begin{lstlisting}[language=Python, firstnumber=11]
# Semantic data slicing detects texts with different surface patterns.
from semslicer.slicer import InteractiveSlicer
%\label{code:data_loading}%data = load_training_data()
%\label{code:slicing_criteria}%criterion = "Muslim"

%\label{code:slicing_config}%slicer = InteractiveSlicer(criterion, data, config={
        'few-shot': True,
        'few-shot-size': 8,
        'instruction-source': 'template',
        'student-model': 'flan-t5-xxl',
        'teacher-model': 'gpt-4-turbo-preview'})

%\label{code:slicing_prompt}%slicer.show_prompt()%\Suppressnumber%
%$\mid$% '''Is the text related to muslim?
%$\mid$% 
%$\mid$% Text: Hardline Muslims are as conservative ...
%$\mid$% Answer: yes
%$\mid$% 
%$\mid$% Text: In fact, Christmas is a pagan festival ...
%$\mid$% Answer: no
%$\mid$% 
%$\mid$% ...
%$\mid$% 
%$\mid$% Text: {text}
%$\mid$% Answer: '''
%\Reactivatenumber%
%\label{code:semantic_slicing}%llm_slicing = slicer.gen_slicing_func() # str -> bool
m_slice = data[data['text'].map(llm_slicing)]
%\label{code:semslicer_result}%m_slice['text'].sample(2)%\Suppressnumber%
%$\mid$% '''The solution to our problem is becoming much more 
%$\mid$%      stricter on immigration, banning those with 
%$\mid$%      religious alliances to Saudi Arabia and its 
%$\mid$%      puppet states. 
%$\mid$%    just look at Turkey over the last 8 years, the 
%$\mid$%      regression under the AKP speaks for itself.'''%\Reactivatenumber%
\end{lstlisting}

\caption{\small Existing practices mostly apply \textit{programmatic} data slicing (cf. Table~\ref{tab:zeno}). For our running example, we can use a simple regex to detect comments with the phrase ``muslim'' or ``islam'' (line~\ref{code:programmatic_slicing}). 
In contrast, \toolname{} supports \textit{semantic} data slicing, by using LLM and generated prompts as slicing functions for any user-provided criteria (line~\ref{code:semantic_slicing}).
The examples here are from the CivilComments dataset~\cite{Borkan2019NuancedMF}---they do not represent the authors' view.}
    \label{fig:slicing_func}
    \vspace{-10pt}
\end{figure}

\begin{table*}[ht]
\centering
\footnotesize
  \begin{tabular}{lllrl}
    \toprule
    \textbf{Slicing Criteria}  & \textbf{Description} &
    \textbf{Examples} & \textbf{Frequency} & \textbf{Mechanism}  \\
    \midrule
    domain & The domain of the text & virology, philosophy  & 65 & Existing annotations \\
    task & The task of the text & data understanding & 27  & Existing annotations \\
    language & The language of the text & french, tamil & 20 & Existing annotations \\
    script & The script of the text & latin, arabic & 6 & Existing annotations \\
    site & The websites an agent needs to access & gitlab, reddit & 6 & Existing annotations \\
    string\_length  & The length of the text (input/output/label) & - & 61 & Simple computation  \\
    num\_steps & The number of steps an agent takes & - & 10 & Simple computation \\
    num\_range & The range of numerical answers & - & 16 & Simple computation \\
    num\_of\_choices  & The number of choices for multi-choice questions & - & 11 & Simple computation \\
    pred\_change\_after\_norm & Whether predictions change after normalization & - & 8 & Simple computation \\
    word\_repetition & The max number of word repetition & - & 3 & Simple computation \\
    output\_shape  & The shape of model output & Yes/No, Valid/Invalid  & 15 & Regex matching \\
    input\_type & The type of input instruction/question & what, how & 11 & Regex matching \\
    is\_X\_library\_used & Whether the input program uses library X  & pandas, numpy & 11 & Regex matching \\
    output\_value  & The value of model output & A, B, no enough info & 5 & Regex matching \\
    has\_function\_calls & Whether the input program contains function calls & - & 4 & Regex matching \\
    topic & The topic of the text & government & 3 & Machine learning (topic modeling) \\
  \bottomrule
\end{tabular}
  \caption{\small We analyzed 20 most popular projects on textual datasets from ZenoHub~\cite{zenohub}, identifying 17 slicing criteria with their descriptions, examples, frequency, and creation methods---almost all are based on existing metadata or can be implemented with simple Python programs.}
  \label{tab:zeno}
    \vspace{-10pt}
\end{table*}

\section{Data Slicing}
\label{sec:background}

\paragraph{Data slicing in ML engineering}
ML engineering is data-centric. 
ML practitioners usually start with data curation to obtain appropriate training and evaluation data;
with LLMs, practitioners may need less data but still need data to develop and test their prompts~\cite{johnnyprompt}.
With curated data, ML practitioners perform data analysis, model training or prompt engineering, and model evaluation (with potential debugging) in multiple iterations.
Afterwards, the model can be deployed, with further model monitoring and updates based on production data. 
All these activities can benefit from \textit{data slicing}.

Data slicing can help \textbf{model debugging}, as illustrated by our running example.
After developers observe model mistakes, conduct error analysis, and formulate hypotheses, data slicing helps them validate the hypotheses on additional relevant examples~\cite{errudite, aisensemaking}.
Once the hypotheses are validated, data slices can be used for further \textbf{model fixing}, via data pre-processing to construct useful features for {model training}, targeted data augmentation, or even as guidance for prompting~\cite{johnnyprompt}.

Data slicing is also useful for fine-grained \textbf{model evaluation},
where multiple behavioral aspects are systematically examined~\cite{ribeiro-etal-2020-beyond, zeno}, as also recognized in the software engineering community~\cite{fse_combi}.
Rather than error-chasing as in model debugging, developers have specific upfront behavioral aspects for model evaluation.
The behavioral aspects can come from the developer's intuition, but can also be elicited from deliberate requirements analysis~\cite[e.g.,][]{improvingdomainspecs, weaver}.
In contrast to traditional benchmarking, where practitioners only examine model accuracy on a single static benchmark,
fine-grained model evaluation can expose nuanced model strengths and weaknesses and help pinpoint areas for {model improvement}.
In our running example of toxicity detection, the developer might want to see if the model treats certain demographic groups systematically unfairly (e.g., regarding race, gender, religion) and use data slicing to identify many concrete examples for each subgroup.

This naturally extends to continuous \textbf{model monitoring},
where developers track model performance on multiple aspects through data slicing~\cite{re2019overton}.
Developers can build regression test suites from data slices~\cite{cain24}, as well as continuously analyze new production data.
From model monitoring, developers can fix degrading aspects when needed, and potentially discover new data patterns (e.g., new hateful slang) for slicing when there is a distribution shift.

In earlier ML engineering phases, data slicing can contribute to \textbf{data curation},
as the insights and data produced from debugging, evaluation, and monitoring can be used to inform which slices are under-performing or under-represented and hence guide what data to curate~\cite{improvingdomainspecs}.

For all these activities, the key motivation for data slicing is to generalize from individual data points to the underlying systematic problems.
This generalization is necessary because model evaluation looks at accuracy in distributions rather than at mistakes for individual inputs -- a single model mistake is not considered a bug~\cite{mlisre}.
This is in stark contrast to traditional software testing, where one single error is considered a bug that can be worth fixing.
Parallels to data slicing, however, do exist in software testing: 
Equivalence class testing~\cite{SEbook} divides the input space into several partitions and creates test cases for each partition, akin to how data slicing partitions datasets into slices; 
strong equivalence class testing explores interaction across these dimensions, which is also explored for data slicing~\cite{slicefinder, fse_combi}.
Despite the parallels, data slices are expected to come from distributions that practitioners care about, rather than from any failing inputs as in software testing.
This key distinction from software testing is why ML engineering benefits from data-centric approaches like data slicing and dedicated innovations like \toolname{}.

\paragraph{Status quo and limitations}
Despite being useful for many activities, (semantic) data slicing is not well-supported.
Data slicing often assumes access to a set of existing features, but relevant features are not always available for users' slicing criteria, limiting developers in practice to create only ``easy'' slices.
We found strong evidence for this by analyzing the 20 most popular projects on ZenoHub~\cite{zenohub}, a platform to share model evaluation results with built-in support for slicing, 
where we observed that developers almost exclusively create \textit{programmatic} slicing functions (e.g., string length, question type) that are easy to implement, as shown in Table~\ref{tab:zeno}.

In contrast, various activities by practitioners and researchers suggest that developers often \textit{want} to conduct \textit{semantic} data slicing that cannot be matched with simple patterns:
For example, \citet{Naik2018StressTE} identified eight different error hypotheses for natural language inference models, 
with half of them being hard to cover with programmatic slicing  (e.g., \textit{``contradicting sentence pairs containing antonyms are hard to classify,''} \textit{``sentence pairs relying on real-world knowledge are hard to classify''}).
Similarly, for machine translation, \citet{karpinska-iyyer-2023-large} report 15 mistranslation hypotheses annotated by humans, most of which are hard to cover with programmatic slicing (e.g., \textit{``translations that change factuality,'' ``translations that are overly literal''}).
\citet{ribeiro-etal-2020-beyond} created a set of 41 functionality tests for three different NLP tasks, with 32 of them hard to implement using programmatic slicing  (e.g., \textit{``author sentiment is more important than that of others''}).

Indeed, precisely because of the difficulty of semantic data slicing, developers resort to synthetic data generation to obtain slices~\cite[e.g.,][]{Naik2018StressTE, ribeiro-etal-2020-beyond} or crowdsourcing to obtain extra annotations for slicing~\cite[e.g.,][]{karpinska-iyyer-2023-large}. 
To bridge this fundamental lack of support for semantic data slicing, \toolname{} employs LLMs as powerful tools to generate semantic slicing functions from any user criteria.

\paragraph{From crowdsourcing to automated semantic slicing}
Recent advancements in LLMs have spurred interest in replacing crowdworkers with LLMs for various tasks~\cite{Wu2023LLMsAW}.
The NLP community has explored using LLMs for data annotations~\cite{zhang-etal-2023-llmaaa}, where researchers manually engineer their prompts for specific tasks.
These annotations can be used as ground-truth labels for model training~\cite{He2023AnnoLLMML, tan2024large}, as noisy labels for weak supervision~\cite{Smith2022LanguageMI, Yu2023AlfredAS}, or as additional clues for model inference~\cite{menon2022visual}.
They have found LLMs often have similar, and sometimes even better performance than crowdworkers~\cite{cegin-etal-2023-chatgpt}.

\toolname{} is motivated by this trend and aims to produce semantic slicing functions that traditionally can only be done by crowdworkers at a high cost.
However, unlike the existing work that relies on manual prompt engineering for specific annotation tasks, \toolname{} is a unifying framework that can be applied to \textit{any} slicing criteria, with a rich set of configuration options for constructing and optimizing the slicing prompt.

\section{Semantic Data Slicing}
\label{sec:approach}

\toolname{} is a framework for \textit{semantic data slicing}, allowing users to create slicing prompts from specified slicing criteria, and use them to generate data slices.
In this section, we will describe (1) the design dimensions of \toolname{} and (2) our system design and implementation details.

\begin{figure}[t]
    \centering
    \includegraphics[width=0.9\linewidth]{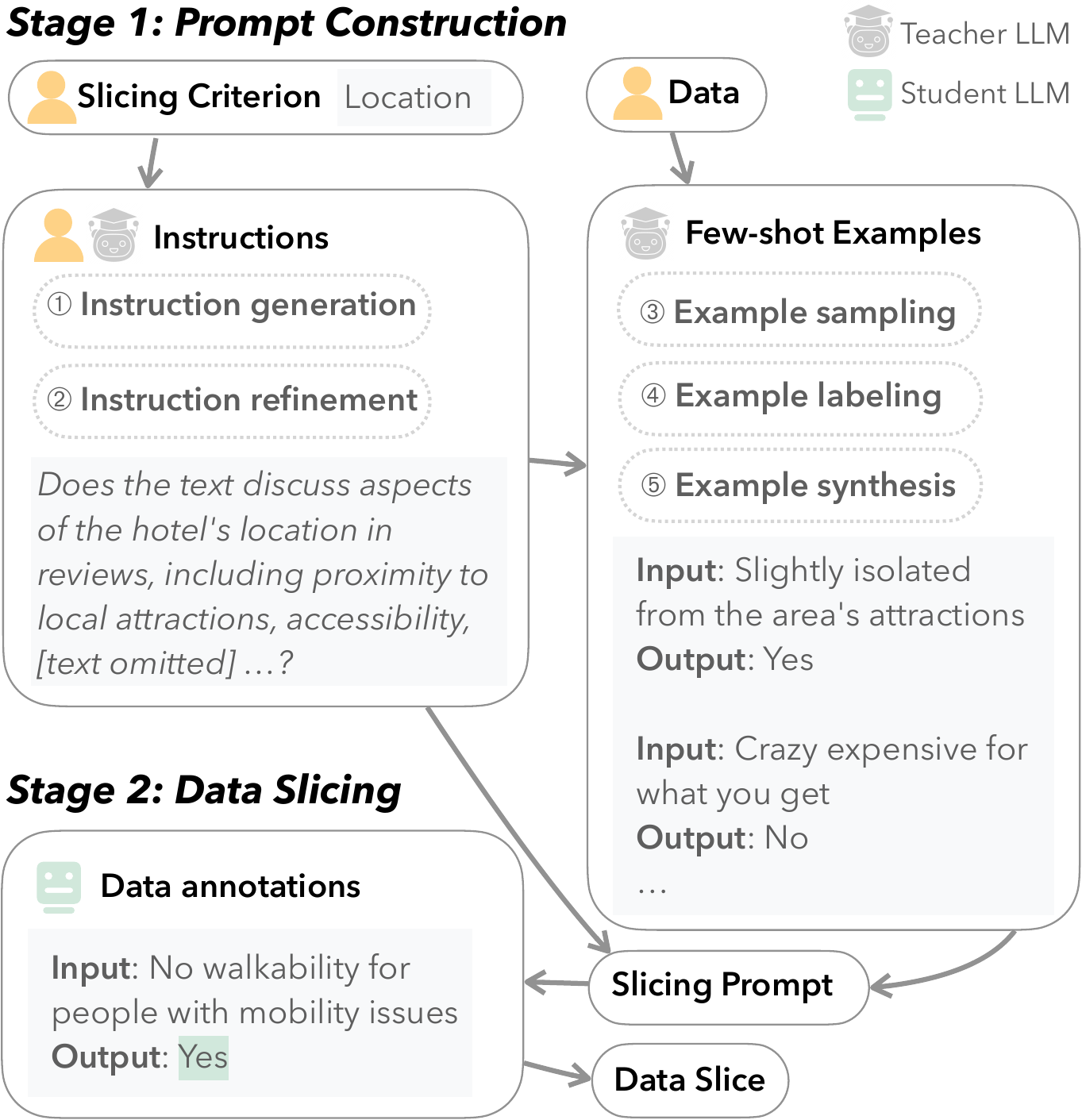}
    \caption{\small 
    \toolname's workflow: The user first specifies a slicing criterion (keywords, descriptions, etc.) and provides a dataset to slice. 
    \toolname{} will \circleone{} construct and \circletwo{} refine a classification instruction from the slicing criterion, optionally with human in the loop. 
    \toolname{} will then \circlethree{} sample and \circlefour{} label few-shot examples, with \circlefive{} synthetic examples generated if needed.
    Finally, \toolname{} uses the produced prompt to annotate the dataset %
    and create the slices.
    }
    \label{fig:overview}
    \vspace{-10pt}
\end{figure}

\subsection{Design Dimensions}
We design \toolname{} considering four dimensions: slicing accuracy needed (\DGone{}), latency expected (\DGfour{}), human effort available (\DGtwo{}), and computational resources available (\DGthree{}).

\subsubsection{Slicing accuracy needed}
Intuitively, we would want higher slicing \DGone{}, as it makes the observed slices more reliable. 
However, as we will demonstrate in Section~\ref{sec:eval-rq3}, moderately accurate slices can also be useful for downstream use cases like model evaluation, as long as they can reliably detect where the model under-performs, making it possible to consider \DGone{} one of the four trade-off dimensions.

\subsubsection{Slicing latency expected}
Our second dimension, \DGfour{}, considers how fast slicing should be.
Depending on the downstream use cases, users can trade off \DGfour{} for other dimensions, or vice versa.
For example, interactive model debugging would expect a lower \DGfour{}, while systematic model evaluation or monitoring can accept a higher \DGfour{}.

\subsubsection{Human effort available}
Our third dimension considers how much \DGtwo{} \toolname{} requires, which also depends on the use cases.
For example, for interactive model debugging, we might want to prioritize lower \DGfour{} over lower \DGtwo{}, as users can put more effort into shaping the slicing function (through prompting) but would expect faster interaction.
In contrast, for model evaluation, we might prioritize \DGtwo{} over \DGthree{} such that evaluation can scale to a larger number of slices.

\subsubsection{Computational resources available}
Our last dimension considers the practical concerns of how much \DGthree{} is available for data slicing.
Low computational cost is important for scaling up \toolname{} in practice.
When resources are limited, users can accept lower \DGone{} to accommodate available resources,
by using a smaller model or having a simpler prompt construction pipeline.

\paragraph{Design trade-off}
There is no optimal design for all four dimensions: We often need to make trade-offs depending on the downstream use cases. 
Our system design aims to allow users to easily trade off along these dimensions,
through using LLMs of different sizes and capabilities (\DGone{} vs. \DGthree{}/\DGfour{}), having a human in the prompt construction loop (\DGone{}/\DGthree{} vs. \DGtwo{}), or having different setups of few-shot examples (\DGone{} vs. \DGthree{}/\DGfour{}).
We will discuss these trade-offs in more detail in Section~\ref{sec:system_design}.

\begin{figure*}
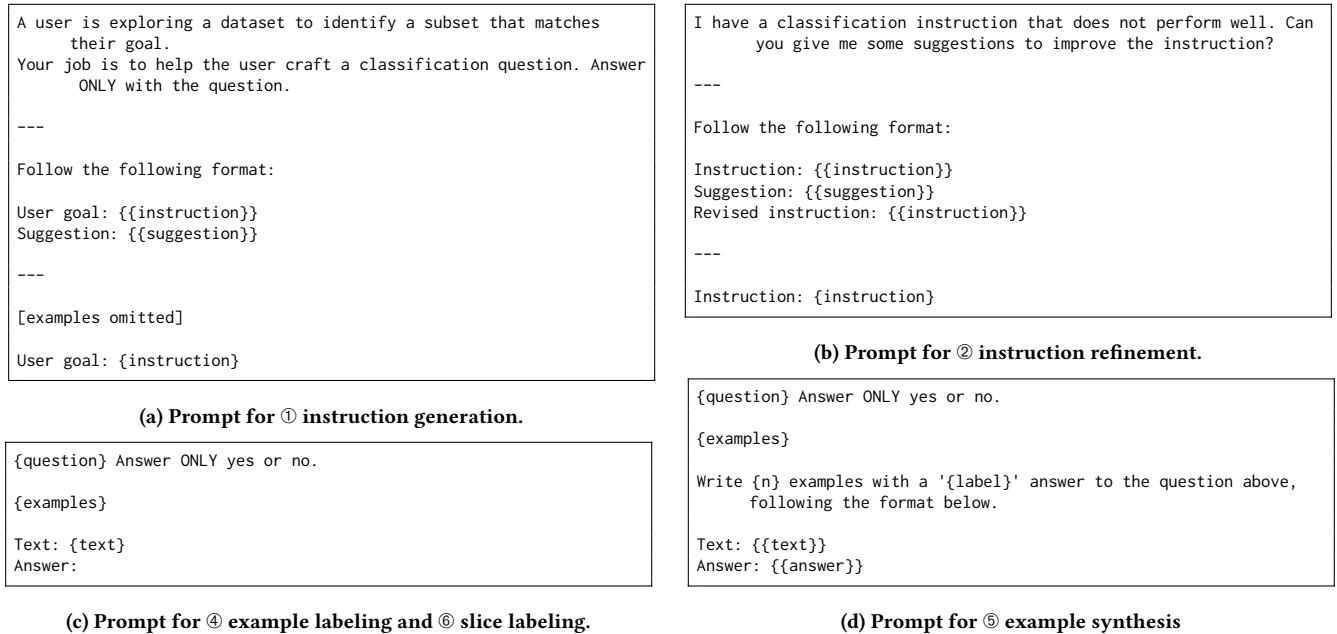

\lstset{
  extendedchars=false,
  basicstyle=\ttfamily,
  columns=fullflexible,
  frame=single,
  breaklines=true,
  postbreak=\mbox{},
}
    \begin{subfigure}[t]{0.47\textwidth}
\footnotesize
        \centering
\begin{lstlisting}
A user is exploring a dataset to identify a subset that matches their goal. 
Your job is to help the user craft a classification question. Answer ONLY with the question.

---

Follow the following format:

User goal: {{instruction}}
Suggestion: {{suggestion}}

---

[examples omitted]

User goal: {instruction}
\end{lstlisting}
        \caption{Prompt for \circleone{} instruction generation.}
    \end{subfigure}%
    \quad\quad
\begin{subfigure}[t]{0.47\textwidth}
\footnotesize
        \centering
\begin{lstlisting}
I have a classification instruction that does not perform well. Can you give me some suggestions to improve the instruction? 

---

Follow the following format:

Instruction: {{instruction}}
Suggestion: {{suggestion}}
Revised instruction: {{instruction}}

---

Instruction: {instruction}
\end{lstlisting}
        \caption{Prompt for \circletwo{} instruction refinement.}
    \end{subfigure}\\
    \begin{subfigure}[t]{0.47\textwidth}
\footnotesize
        \centering
        \vspace{4pt}
\begin{lstlisting}
{question} Answer ONLY yes or no.

{examples}

Text: {text}
Answer: 
\end{lstlisting}
        \caption{Prompt for \circlefour{} example labeling and \circlesix{} slice labeling.}
    \end{subfigure}
    \quad\quad    
\begin{subfigure}[t]{0.47\textwidth}
\vspace{-20pt}
\footnotesize
        \centering
\begin{lstlisting}
{question} Answer ONLY yes or no.

{examples}

Write {n} examples with a '{label}' answer to the question above, following the format below.

Text: {{text}}
Answer: {{answer}}
\end{lstlisting}
        \caption{Prompt for \circlefive{} example synthesis }
    \end{subfigure}

    \caption{Prompt templates used in \toolname{}.}
    \label{fig:prompts}
    \vspace{-10pt}
\end{figure*}

\subsection{System Design}
\label{sec:system_design}

\paragraph{System overview}
At a high level, \toolname{} runs in two stages (as depicted in Figure~\ref{fig:overview}).
In the first \textit{prompt construction} stage, \toolname{} constructs a slicing prompt for a user-provided slicing criterion. 
\toolname{} will first \circleone{} generate and \circletwo{} refine a classification instruction, and then \circlethree{} sample, \circlefour{} label, and optionally \circlefive{} synthesize few-shot examples. 
Depending on the configuration, some steps can be skipped or customized.
In the second \textit{data slicing} stage, \toolname{} will annotate the dataset using the constructed prompt and produce a corresponding slice.

\paragraph{Slicing function components}
The final slicing function produced by \toolname{} will consist of three components: The model (and inference settings) used, the instructions specified, and the few-shot examples provided (see Figure~\ref{fig:slicing_func}, bottom). 
All three components can greatly impact slicing \DGone{} and \DGfour{} and need different levels of \DGtwo{} and \DGthree{}.

\begin{itemize}
    \item \textbf{Model}: Model is the most critical component as LLM capabilities can greatly impact task \DGone{}.
    In \toolname{}, we leverage LLMs in many different steps, and for each step, we need to make trade-offs between \DGone{} and \DGthree{}/\DGfour{} to choose an appropriate model.
    \item \textbf{Instructions}: Instructions state what a model should do, often in the form of a classification question in \toolname{} (see example instruction in Figure~\ref{fig:overview}). LLMs are shown to have strong instruction-following capabilities~\cite{ouyang2022training}, but the quality of responses depends on the exact instructions provided~\cite{srivastava2022beyond}.
    \toolname{} supports using LLMs to help generate and refine instructions, with a human in the loop, inspired by recent advancements in automated prompt engineering~\cite{zhou2023large, pryzant-etal-2023-automatic}.
    Turning this option on can improve \DGone{} at the cost of more \DGtwo{} or \DGthree{}.
    \item \textbf{Few-shot examples}: Few-shot examples are demonstrations that show how a model should respond to a task input (see example input/output pairs in Figure~\ref{fig:overview}). They also have a strong impact on LLM performance~\cite{NEURIPS2020_1457c0d6, lu-etal-2022-fantastically}. 
    \toolname{} supports a rich set of tools to construct few-shot examples, such as different sampling strategies, labeling strategies, and synthetic input generation.
    Users can make trade-offs between \DGone{} and \DGthree{}/\DGfour{} by choosing the right set of configurations. 
\end{itemize}

Next, we detail our design choices, rationale, and implementations for these three components in \toolname{}.

\subsubsection{Model}
We design \toolname{} such that different steps and stages can flexibly leverage different LLMs, allowing trade-offs between \DGone{} and \DGthree{}/\DGfour{}.
The general rationale behind our design is that a more powerful model can produce higher-quality outputs, but would increase the compute needed, incurring higher cost (and latency), while a less powerful model would be cheaper yet less accurate.
Users should have the agency to decide which model fits best for their use case, according to their specific constraints.

In \toolname{}, LLMs are used in almost  every step:
In \circleone{} \textbf{instruction generation}, \circletwo{} \textbf{instruction refinement}, and \circlefive{} \textbf{example synthesis}, 
we need a model with human-like creativity and capabilities to generate and refine instructions, as well as synthesize new examples.
In \circlefour{} \textbf{example labeling}, we need a model with high-quality classification capabilities for small-scale labeling.
In \textbf{data slicing}, we need a model with good-quality classification capabilities for larger-scale labeling.

\paragraph{Implementation}
In our evaluation, we use \texttt{gpt-4-turbo-preview} for steps \circleone{}\circletwo{}\circlefive{} (temperature=1) and step \circlefour{} (temperature=0),\footnote{Temperature is a parameter commonly used to control LLM outputs. A higher temperature makes the model more ``creative,'' while a lower temperature makes the model more predictable. We use temperature=1 for creativity as recommended~\cite{textgenAPI}, while temperature=0 is a common setup for classification tasks.}
as GPT-4 is known to be among the strongest LLMs available at the time of our implementation but also incurs higher cost and latency. 
These steps are all in the prompt construction stage and hence executed less frequently per slice.

We use \texttt{flan-t5-xxl} for the second stage (temperature=0), as the model has strong classification capabilities~\cite{chung2022scaling} and is of a moderate size with 11B parameters, hence lower cost and latency,
which is crucial in the data slicing stage as the entire dataset is annotated. 
We further use 8-bit quantization to speed up model inference.

\subsubsection{Instructions}
We design \toolname{} to allow more LLM-based automated prompt engineering (\DGthree{}), with a flexible human-AI collaboration mechanism (\DGtwo{}) to improve instruction quality (\DGone{}).
This is achieved through two steps: \circleone{} \textbf{instruction generation}, to produce a good initial instruction, and \circletwo{} \textbf{instruction refinement}, to further refine the produced instruction.

The first step aims to produce an initial instruction from a user-provided slicing criterion.
Depending on what users provide, \toolname{} provides options for (1) template-based construction and (2) LLM-based generation:
\begin{enumerate}
    \item \textit{Template}: When users provide a criterion in simple phrases, \toolname{} can use the generic template \textit{``Is the text related to \{concept\}?''} to translate a criterion into an instruction. 
    We observe this simple template works very well for many slicing tasks (cf. Section~\ref{sec:intrinsic_eval}).
    \item \textit{LLM-generated}: When users provide more detailed descriptions, \toolname{} can leverage an LLM, which offers strong information processing capabilities, to generate an initial instruction.
\end{enumerate}

The produced instruction provides a good starting point, which can be further refined in the second step.
This can be achieved through (1) human post-editing, (2) LLM-based refinement, or (3) a mixture of both.
For LLM-based refinement, our design is inspired by Self-Refine~\cite{madaan2024self}, where we ask an LLM to provide suggestions and revisions on a to-be-revised instruction.
This refinement step can increase \DGone{} at the cost of more \DGtwo{} or \DGthree{}.

\paragraph{Implementation}
We prompt an LLM for both \circleone{} \textbf{instruction generation} and \circletwo{} \textbf{instruction refinement} (Figure~\ref{fig:prompts}).

When using the instruction in slice labeling, \toolname{} deliberately instructs the model to only produce a label token (Figure~\ref{fig:prompts}), without any intermediate reasoning steps (e.g., as in chain-of-thought prompting~\cite{wang2022self}).
This design is based on our observation that the intermediate steps sometimes help \DGone{}, yet incur high \DGfour{}, which scales linearly with the number of output tokens
(\DGone{} vs. \DGfour).

\subsubsection{Few-shot examples}
The design of \toolname{} automates the construction of few-shot examples, which demonstrate how a model should label a slice.
Users can make trade-offs between \DGone{} and \DGthree{}/\DGfour{} by controlling different options, including (1) whether they need few-shot examples at all, (2) how many examples they need, (3) where the examples are from, and (4) how to label the examples.
The more examples they use in the slicing prompt, the higher \DGthree{} cost and \DGfour{}.

\looseness=-1
As a first step to construct few-shot examples, \toolname{} needs to curate a set of example inputs.
Assuming access to user-provided data, \toolname{} provides different \circlethree{} \textbf{sampling strategies}: 
(1) random sampling and (2) diversity sampling. 
We support diversity sampling as we observe that random sampling can miss examples from smaller slices, which biases the second stage of \textbf{data slicing}.

With the sampled example inputs, \toolname{} next labels the examples with the generated slicing instruction.
In this step, we need to bootstrap from zero-shot labeling (i.e., without any examples), as there are usually no existing labels on a user's slicing criteria, which can be arbitrary. 
\toolname{} supports different \circlefour{} \textbf{labeling strategies}, (1) student-label and (2) teacher-label, to accommodate different trade-off decisions (\DGone{} vs. \DGthree{}).
A teacher model is a stronger LLM (e.g., GPT-4), whose labels are usually more accurate but also more expensive, while a student model is a smaller LLM that is less accurate but also cheaper.
Using teacher-labeled examples can effectively distill a teacher model's knowledge about some particular slicing criterion to a student model.

In the last step, \toolname{} creates \circlefive{} \textbf{optional synthetic examples} to balance the in-slice and out-of-slice examples, as sometimes sampled examples can be extremely imbalanced for small slices, leaving no in-slice demonstrations for data slicing.
To create synthetic examples that look similar to the real dataset, \toolname{} queries an {LLM} to write extra examples of the underrepresented label, \textit{given sampled examples}. 
The given examples can condition the LLM on what inputs (style, content, etc.) it should generate.

\paragraph{Implementation}
We implement diversity sampling by (1) vectorizing user-provided data using SentenceTransformer embeddings~\cite{reimers-2019-sentence-bert} and (2) clustering the data into $N$ clusters with KMeans and selecting one example from each cluster.
This strategy produces semantically different clusters and hence diverse examples.

In our evaluation, we set the number of few-shot examples to 8, following existing work~\cite{chung2022scaling}.
For \circlefour{} example labeling, we experiment with both a teacher model (\texttt{gpt-4-turbo-preview}) and a student model (\texttt{flan-t5-xxl}).
For \circlefive{} example synthesis, we use \texttt{gpt-4-turbo-preview}  (Figure~\ref{fig:prompts}).

\subsubsection{Data slicing}
With the slicing prompt constructed, the final {data slicing} step applies the prompt using a student model---we use \texttt{flan-t5-xxl} in our evaluation.
This step produces slicing annotations for the entire dataset.

\begin{table*}[ht]
  \centering
  \small
  \begin{tabular}{llllllr}
    \toprule
    \addlinespace
    {\textbf{Method}}&\multicolumn{5}{c}{\textbf{Configurations}} & \multirow{1}{*}{\raggedleft\textbf{Cost/Slice}} \\    
            &\multicolumn{3}{c}{\textbf{Few-shot (fs) Examples}}
            &\multicolumn{2}{c}{\textbf{Instructions}} &
\\    
    \cmidrule(lr){2-4}
    \cmidrule(lr){5-6}
            &{Input source} & {Input sampler} & {Output labeler} &{Source} &{Refinement} 

\\
    \midrule
    M$_{\text{zero-shot}}$ & - (zero-shot, zs) & - & - & template & - & \$0.26 \\
    M$_{\text{few-shot}}$ & provided & random & student & template & - & \$1.68 \\
    M$_{\text{fs-div}}$ & provided & diversity (div) & student & template & - & \$1.68 \\
    M$_{\text{fs-teacher}}$ & provided & diversity & teacher & template & - & \$1.69 \\
    M$_{\text{fs-syn}}$ & provided + synthesized (syn) & diversity + balanced & teacher &  template & - & \$1.70 \\
    M$_{\text{zs-model}}$ & - & - & - & model & - & \$0.26  \\
    M$_{\text{zs-tmodel}}$ & - & - & - &  template (t) & model & \$0.26 \\
    M$_{\text{zs-hai}}$ & - & - & - &  human + template & human + model (hai) & \$8.13 \\
    M$_{\text{fs-hai}}$ & provided & diversity & teacher &  human + template & human + model (hai)  & \$9.56 \\
    Crowdworker & - & - & - &  human & - & \$432.00 \\
  \bottomrule
\end{tabular}
  \caption{\small Experiment configurations: We selected 9 representative configurations of \toolname{}, following a fractional factorial design~\cite{bailey2008design}.
  We estimate that \toolname{} costs from \$0.26 to \$9.56 to generate one slice from a dataset of 6k examples (CivilComments).
  In contrast, using crowdworkers for the same task would cost \$432.00, which is 44x to 1661x more than \toolname{} (see the appendix for how we estimate cost).
 }
  \label{tab:exp_config}
    \vspace{-10pt}
\end{table*}

\subsection{System Interface}
\toolname{} is packaged as a Python library. 
As shown in Figure~\ref{fig:slicing_func}, a user first provides a dataset to slice (line~\ref{code:data_loading}) and a slicing criterion (line~\ref{code:slicing_criteria}).
Next, the user specifies what slicing function they want to generate with the desired configurations (line~\ref{code:slicing_config}), with fine-grained control on each component of \toolname{}.
\toolname{} will generate a slicing prompt, following the workflow summarized in Figure~\ref{fig:overview}, and produce a slicing function (line~\ref{code:semantic_slicing}), which can be further applied on any datasets (line~\ref{code:semslicer_result}).
The entire process can be easily batched and extended to multiple slices.

The above example demonstrates how \toolname{} can support workflows with minimal \DGtwo{}.
However, as we have discussed, humans can easily provide more guidance by (1) providing more detailed slicing instructions instead of simple keywords (line~\ref{code:slicing_criteria}), (2) post-editing the generated prompt (line~\ref{code:slicing_prompt}), or (3) iterating the entire process for a few rounds.

\section{Evaluation}

\label{sec:intrinsic_eval}
First, to demonstrate feasibility and practicality, we evaluate our method's ability to produce accurate slices and its associated cost across \textit{different} system configurations:
\begin{itemize}
    \item \textbf{RQ1}: How accurate are \toolname{}'s predicted slices across different configurations?
    \item \textbf{RQ2}: How much cost/latency does it take to produce the slices across different configurations? 
\end{itemize}

Next, to demonstrate usefulness for downstream usage, we evaluate our method's ability to assist model evaluation:
\begin{itemize}
    \item \textbf{RQ3}: How useful is \toolname{} for model evaluation?
\end{itemize}

\begin{table*}[t]
 \centering
 \footnotesize
  \begin{tabular}{llrrrrrrrrrrll}
    \toprule
    \addlinespace
    {\textbf{Dataset}} & {\textbf{Slice}} & {\textbf{Frac. (\%)}} 
    & \multicolumn{9}{c}{\textbf{Slicing F1 score (\%)}}   
    & \multicolumn{2}{c}{\textbf{Task perf. (\%)}} \\    
    \cmidrule(lr){4-12}
     \cmidrule(lr){13-14}
        & & & 

    M$_{\text{zero-shot}}$ & 
    M$_{\text{few-shot}}$ & 
    M$_{\text{fs-div}}$ & 
    M$_{\text{fs-teacher}}$ & 
    M$_{\text{fs-syn}}$ & 
    M$_{\text{zs-model}}$ & 
    M$_{\text{zs-tmodel}}$ & 
    M$_{\text{zs-hai}}$ & 
    M$_{\text{fs-hai}}$ &
    gold & 
    pred.  
        \\
    \midrule
     \multirow{8}{*}{\textbf{HateCheck}} 
    & women & 13.7 & 90.7 & \markgood 92.7 & \markgood 92.6 & \markgood 92.8 & 91.7 & 92.3 & \markpoor 89.5 & 90.5 & \markbest 94.6 & 79.4 &  82.6   \\
    & trans & 12.4 & 83.6 & 79.5 & 78.3 & \markgood 95.1 & 94.7 & \markgood 95.1 & \markpoor 75.7 & 90.3 & \markbest 96.5  & 53.3** & 52.7** \\
    & gay & 14.8 & \markpoor 68.3 & 69.0 & 69.5 & 69.7 & 68.8 & \markgood 85.0 & 69.2 & \markbest 88.4 & 71.7  & 70.6 & 63.8** \\
    & black  & 12.9 & \markgood 95.2 &  \markgood 95.4 & \markgood 95.1 & \markgood 95.1 & \markbest 95.7 & \markgood 95.2 & \markpoor 86.3 & 94.9 & \markgood 95.1  & 70.7  & 71.6 \\
    & disabled & 13.0 & 68.1 & \markgood 77.9 & \markpoor 67.3 & \markpoor 67.4 & \markgood 81.6 & 71.4 & 69.2 & \markbest 89.4 &  73.3  & 47.3** & 48.7** \\
    & muslim & 13.0 & \markbest 97.7 & \markgood 97.1 & 95.8 & 95.4 & \markpoor 93.9 & \markgood 97.1 & 95.8 & \markgood 97.2 & 95.4  &  78.3 & 79.9\\
    & immigrant & 12.4 & 80.0 & 77.2 & 66.8 & 82.2 & 68.5 & \markgood 94.5 & \markpoor 21.9 & \markbest 94.9 &  \markgood 93.5 & 71.5 & 74.3 \\
    & avg & - &  83.4 &84.1 &  80.8 & {85.4} &  85.0 & \markgood {90.1} & \markpoor 72.5 & \markbest 92.2 & \markgood 88.6 & 69.4 & 69.4\\
    \midrule
     \multirow{7}{*}{\textbf{AdaTest}} 
     & room & 24.9 & 59.3 & 84.9 & \markgood 88.2 & \markbest 89.6 & 87.4 & \markpoor 15.4 & 70.9 & 84.7 & \markgood 88.6 &66.0&  65.3\\
     & location & 16.9 & 49.1 & 74.1 & 73.6 & \markgood 82.7 & 80.5 & \markpoor 28.6 & 48.5 & \markgood 82.4 & \markbest 87.9 &62.5& 58.1 \\
     & price & 18.5 & \markgood 95.9 & \markgood 95.8 & \markgood 95.8 & 94.3 & 94.3 & \markgood 95.8 & \markpoor 24.4 & \markbest 97.2 & \markgood 95.8 &65.7& 62.9\\
     & restaurant & 17.5 & \markpoor 57.1 & 76.5 & 65.1 & \markgood 78.1 & 62.4 & 59.2 & 58.3 & \markgood 78.6 & \markbest 93.8 &63.6& 74.2 \\
     & service & 15.9 & \markpoor 37.3 & 49.2 & 49.6 & \markgood 54.1 & 48.8 & 45.8 & 43.3 & \markbest 59.4 & \markgood 55.0  &  80.0& 76.5\\
     & pool & 6.3 & \markgood 85.7 & \markbest 90.9 & \markbest 90.9 & \markpoor 73.7 & \markbest 90.9 & 80.0 & \markpoor 73.7 & \markbest 90.9  & \markbest 90.9 &50.0& 57.1\\
    & avg & - & 64.1 & {78.6} & 77.2 & {78.7} & 77.4 & \markpoor 54.1 & \markpoor 53.2 & \markgood 82.2 & \markbest 85.3 &66.1& 66.1 \\
    \midrule
     \multirow{6}{*}{\textbf{Amazon}}
     & book & 61.4 & 95.9 & \markgood 96.5 & \markgood 96.3 & \markbest 97.0 & \markgood 96.7 & 95.9 & \markgood 96.6 &  \markpoor 94.7 & 95.1 &68.7 & 68.7 \\
     & movie & 7.0 & \markpoor 53.1 & 55.7 & 55.9 & \markgood 59.1 & 56.3 & \markgood 58.1 & 55.0 & 56.0 & \markbest 81.1 & 69.6& 66.5\\
     & home\&kitchen & 6.3 & 52.1 & \markpoor 47.2 & 48.5 & 48.3 & 45.8 & 50.3 & \markbest 59.9 & \markgood 59.7 & 48.1 &73.3& 69.1 \\
     & electronics & 5.8 & \markpoor 52.1 & 54.3 & 57.8 & \markgood 60.0 & 56.6 & \markpoor 52.1 & \markgood 62.8 & 61.7 & \markbest 64.9 &74.7& 72.4\\
     & clothing & 5.4 & \markpoor 63.8 & 72.0 & 65.9 & 72.8 & 66.3 & 66.1 &  \markgood 75.1 & \markgood 74.8 & \markbest 79.1  &71.8& 72.5\\
    & avg & - & \markpoor 63.4 & 65.2 & 64.9 & {67.4} & 64.3 & 64.5 & \markgood 69.9 & \markgood 69.4 & \markbest 73.7  &69.8& 69.8\\
    \midrule
     \multirow{9}{*}{\textbf{CivilComments}}
    & male & 2.5 & 12.3 & \markgood 30.6 & \markbest 32.3 & \markgood 29.8 & \markpoor 5.1 & 10.6 & 25.3 & 9.3 & 10.9 & 89.3 & 90.7 \\
    & female & 2.9 & 30.1 & \markgood 40.9 & \markgood 40.9 & \markbest 42.5 & \markpoor 17.3 & 30.9 & 33.2 & 31.5 & 36.3 & 87.3** & 89.4* \\
    & homosexual & 0.5 & 31.3 & 38.8 & \markgood 43.5 &  40.9 & 40.6 & 35.2 & \markpoor 25.3 & \markgood 41.8 & \markbest 50.6  & 82.8* & 87.8* \\
    & christian  & 2.3 & 53.1 & \markbest 59.2 & \markgood 56.1 & \markpoor 51.0 & 55.3 & 54.3 & \markgood 57.0 & 51.3 & 51.5 & 91.9 & 93.8 \\
    & jewish &  0.5 & 50.0 & 55.0 & \markgood 69.7 & \markgood 69.1 & 62.3 & 50.0 & \markpoor 48.4 & 49.5 & \markbest 85.7 & 92.6 & 96.4 \\
    & muslim & 1.2 & \markpoor 60.9 & 71.3 & 69.3 & 73.1 & \markgood 79.2 & 70.7 & 71.1 & \markgood 78.2 & \markbest 82.0 & 87.3* & 88.5* \\
    & black &  0.8 & 43.2 & 54.3 & 64.5 & \markgood 69.1 & \markgood 66.1 & \markpoor 31.0 & 44.4 & 44.7 & \markbest 71.0 & 75.0** & 79.0** \\
    & white & 1.4 & 73.0 & 81.5 & \markgood 83.6 & 78.7 & \markbest 85.5 & 67.5 & \markpoor 32.5 & 78.1 & \markgood 85.2 & 81.4** & 78.1** \\
    & avg & - & 44.2 & 54.0 & \markgood {57.5} & \markgood {56.8} & 51.4 & 43.8 & \markpoor 42.1 & 48.1 & \markbest 59.2 & 93.5 & 93.5 \\
    \midrule
    - & avg &  - &  63.0 & 69.9 & 69.7 & \markgood 71.6 & 68.9 &  62.6 & \markpoor58.2 & \markgood 71.9 & \markbest 75.9 & - & -\\
     \bottomrule
      \multicolumn{11}{l}{\fcolorbox{black}{darkgreen}{} best \quad \fcolorbox{black}{lightgreen}{} good \quad \fcolorbox{black}{rose}{} poor} & \multicolumn{3}{r}{\scriptsize{$^{**}p<0.01,\quad ^{*}p<0.05$}}
\end{tabular}
  \caption{\small We collected 4 datasets and 26 slices for our evaluation. 
  These slices represent a 0.5\% to 61.4\% fraction of the entire dataset.
  We found that \toolname{} achieves an average F1 score of up to 75.9\%  (achieved by M$_{\text{fs-hai}}$), with significant improvement from few-shot examples (+4.0\% vs. M$_{\text{zs-hai}}$) and human interventions (+4.3\% vs. M$_{\text{fs-teacher}}$) (RQ1).
  We also found that \toolname{} can recover 7 out of 7 slices that have significantly lower downstream task performance, with only one false positive (RQ3).}
 \label{tab:result}
    \vspace{-10pt}
\end{table*}

\subsection{Experiment Setup}
\label{sec:setup}

\paragraph{Datasets}
To evaluate our method, we collect existing datasets with \textit{ground truth} slices, as there are no existing benchmarks on semantic data slicing.
We look at three different strategies to collect suitable datasets:

\begin{itemize}
    \item \textit{Human annotations}. 
    Sometimes dataset curators ask humans (usually crowdworkers) to annotate existing datasets with \textit{additional} attributes of input texts (e.g., ethnicity groups referenced in texts). 
    These attributes make plausible slices that are hard to obtain without human annotations.
    These slices are the most realistic (though potentially noisy due to crowdworker mistakes and subjectivity~\cite{Aroyo2015TruthIA}), but they are challenging to find due to the high cost of human annotation.

    We use the existing {CivilComments}~\cite{Borkan2019NuancedMF} dataset---the task is toxicity detection, but in addition to the toxicity label, the dataset has been annotated by crowdworkers with additional attributes regarding referenced demographic groups on five categories (gender, sexual orientation, religion, race, and disability) with 23 concrete attributes (e.g., Black, Christian). 
    We derive slices from these additional attributes.
    To reduce experiment cost, we randomly sampled 6000 examples and used the 8 largest slices.

    \item \textit{Synthetic data}. 
    For model testing and evaluation~\cite{ribeiro-etal-2020-beyond}, developers often create synthetic test data for different subgroups of the target population.
    We can treat data from each subgroup as a slice. 
    These slices are accurate by construction, as they are created specifically for each subgroup, but can be less realistic due to their synthetic nature. 
    
    We use two datasets created with this strategy, Hate\-Check~\cite{rottger-etal-2021-hatecheck} and {AdaTest}~\cite{ribeiro-lundberg-2022-adaptive}.
    The first is a hate speech test suite (n=3728) with comments  for 7 different subgroups (e.g., women, trans). 
    The second is a sentiment analysis test suite (n=196) with  data for 6 aspects (e.g., price, location) of a hotel review. 
    
    \item \textit{Metadata}. 
    Many datasets come with metadata produced as part of data collection, such as tags on QA websites from which data was scraped. 
    We can create slices from such metadata, but these slices may be less accurate as different categories (e.g., clothes vs. sports) might overlap semantically, leaving some examples missing from the slices.
    
     We use the existing Amazon~\cite{blitzer-etal-2007-biographies} sentiment analysis dataset, which in addition to the sentiment label, contains metadata about what product category was reviewed (e.g., books, electronics). 
     To reduce experiment cost, we randomly sampled 6000 examples and used the 5 largest slices.
\end{itemize}

To summarize, we selected four datasets with 26 ground-truth slices---see Table~\ref{tab:result} for their names and proportions.
These datasets cover different uses of slicing: slicing on sensitive attributes for fairness (HateCheck and CivilComments) and slicing on topics/domains for fine-grained model evaluation (AdaTest and Amazon).
We collected all slicing criteria (the inputs to \toolname{}) directly from the datasets, or from the descriptions in the associated papers.

\paragraph{Configurations}
To understand how different configuration options impact the accuracy and cost of \toolname{},
we selected and analyzed 9 representative configurations of \toolname{}, following a fractional factorial design~\cite{bailey2008design} to cover all values in each dimension (as shown in Table~\ref{tab:exp_config}). 
These configurations cover whether to use few-shot examples (zero-shot vs. few-shot), how few-shot inputs are collected (provided vs. synthetic), sampled (random vs. diversity), and labeled (student vs. teacher), as well as how instructions are generated (template vs. model vs. human+template) and refined (model vs. human+model).

For the configurations with human interventions (M$_{\text{zs-hai}}$ and M$_{\text{fs-hai}}$), we have one of the authors (1) write down slicing descriptions for the model and (2) post-edit the model-refined instructions.

{
Note that here we deliberately compare different configurations of} \toolname{}{, instead of comparing it to a baseline using regular expression.
This is because programmatic slicing performs too poorly to make a meaningful baseline:
For example, our analysis shows that direct keyword matching on the literal keyword ``location'' yields low recall (0.059), and a refined regex approach (r``location|walk|far from|close to|neighborhood|near'') only improves recall to 0.65. 
Even such a mediocre performance already requires extensive user labor---we crafted the regex after carefully inspecting 30\% of examples from the ground-truth ``location'' slice, which results in overfitting and demonstrates the limitations of rule-based systems. Including such an obviously poor baseline would not have added meaningful value to the evaluation.
}

\paragraph{Threats to Validity}
Despite best efforts, the datasets we used are not perfect, with noisy labels or less realistic inputs as explained above.
The human-in-the-loop configurations are also limited by the authors' prompt writing expertise.
As is common in these kinds of studies in real-world settings, our human-subject case study trades off lower internal validity for higher external validity, in a context with limited control over the setting and limited ability to perform repeated independent observations.
Generalizations beyond our evaluation results should be done with care.

\toolname{} uses LLMs at multiple steps. While LLMs can be non-deterministic and make the results less reliable, we deliberately used temperature=0 setting for the data slicing step. 
That is, repeatedly sending the same prompt to the same LLM will produce the same answer. We also used an open-source LLM, making this step fully reproducible.
For instructions and synthetic examples in prompt construction, the variances from LLM non\--determinism are smoothed over as we average results from 26 slices.
\footnote{{
To better understand the evaluation variance for individual slices, we ran M$_{\text{zs-model}}$ five times on seven HateCheck slices.
We observed that the standard deviation ranges from 0.0017 to 0.061, with maximal F1 differences up to 14\%. After averaging, the standard deviation is 0.013, and the maximal F1 difference is only 3.5\%, supporting our point that the averaged results reduce evaluation variances.}}

\subsection{RQ1: Slicing Accuracy}
\paragraph{Setup}
We measure \textit{slicing accuracy} for each slice, that is, how well the slices generated by \toolname{} correspond to the ground-truth slices in the dataset.
We measure slicing accuracy using F1-score, an established metric for imbalanced datasets, as slices often represent a small fraction of the entire dataset (ranging from 0.5\% to 61.4\% in our evaluation, see Table~\ref{tab:result}).
We additionally compute the average F1-score across all slices from each dataset, as well as average F1-score overall.

\paragraph{Result: \toolname{} produces accurate annotations with 75.9\% average F1-score across all slices in the best configuration.}

\toolname{} produced the most accurate slices with 75.6\% average F1-score in configuration M$_{\text{fs-hai}}$ with few-shot examples and human+model refinement, as shown in Table~\ref{tab:result}.
For the configurations without human intervention, we found M$_{\text{fs-teacher}}$ has the highest average F1-score of 71.6\%.
As discussed earlier, \toolname{} does not need perfect accuracy to generate useful slices---our results for RQ3 will show that the current level of accuracy can already reliably detect under-performing slices.

\begin{table}[t]
 \centering
 \footnotesize
  \begin{tabular}{lrrrr}
    \toprule
    \textbf{Dataset}
        & 
    \multicolumn{2}{c}{\textbf{Annotations/sec.}} & 
    \multicolumn{2}{c}{\textbf{Tokens/annotation}}  \\    
    \cmidrule(lr){2-3} 
    \cmidrule(lr){4-5}
        & 
    zero-shot & 
    few-shot & 
    zero-shot & 
    few-shot 
        \\
    \midrule
     {\textbf{HateCheck}} & 12.25 & 10.99 & 32.12 & 178.73 \\
     
     {\textbf{AdaTest}} & 14.56 & 10.84 & 27.45 & 111.79 \\
    
     {\textbf{Amazon}} & 14.79 & 6.01 & 131.40 & 648.90 \\
    
    {\textbf{CivilComments}}  & 14.46 & 6.45 & 93.17 & 626.54\\\addlinespace
    
    \textbf{Average} & 14.07 & 7.40 & 92.47 & 523.79  \\
     \bottomrule
\end{tabular}
  \caption{\small \toolname{} can produce slices at a fast speed: 
  Depending on the exact setup (zero-shot vs. few-shot) and dataset characteristics (tokens per annotation), \toolname{} can annotate from 6 to 14 examples per second.
  As an example, annotating the {\textbf{CivilComments}} dataset (n=6k) takes around 15.5 minutes, using 2 A6000 GPUs.}
 \label{tab:compute}
    \vspace{-20pt}
\end{table}

\paragraph{Result: Few-shot prompting improves accuracy by 7\% on average.}
Comparing few-shot configurations with zero-shot,  we found few-shot prompting improves performance by 7\% on average (Table~\ref{tab:result}).
Among different few-shot setups, we found teacher labeling improves performance by 1.9\% (M$_{\text{fs-teacher}}$ vs. M$_{\text{fs-div}}$), 
but observed a limited impact from sampling methods (M$_{\text{few-shot}}$ vs. M$_{\text{fs-div}}$) and a negative impact from synthetic examples (M$_{\text{fs-teacher}}$ vs. M$_{\text{fs-syn}}$).
We hypothesize that this is because the generated synthetic examples are still too unrealistic to help later slice labeling.

\paragraph{Result: Human intervention significantly improves accuracy.}
We observe that LLM-based instruction generation and refinement have an unstable impact compared to simple templates: They sometimes generate good instructions that improve accuracy a lot (e.g., M$_{\text{zs-model}}$ for HateCheck) but can also generate bad instructions that hurt accuracy (e.g., M$_{\text{zs-tmodel}}$ for HateCheck).
{However, with human interventions, generated instructions} significantly improve accuracy, with an 8.9\% increase for zero-shot (M$_{\text{zero-shot}}$ vs. M$_{\text{zs-hai}}$), and a 4.3\% increase for few-shot (M$_{\text{fs-teacher}}$ vs. M$_{\text{fs-hai}}$). 
{This shows that human feedback is particularly useful to guide instruction optimization in the right direction.}

\paragraph{Discussion}
\label{sec:crowd-acc}
Because we noticed that \toolname{} performs poorly on some slices of the CivilComments dataset, especially due to low precision,
we investigated the reasons for this behavior.
We sampled 40 false positive examples from four slices with low precision scores (\textit{male}, \textit{female}, \textit{homosexual}, \textit{christian}) under the M$_{\text{fs-teacher}}$ setting and manually inspected the examples:
Surprisingly, we found that 24 out of 40 examples actually cover nuanced patterns that (we argue) should be in the slices, yet were missed by human annotators. 
These patterns are often related to the slicing criterion in a less direct way (e.g., ``Pride'' for \textit{homosexual}; ``scripture'', ``church'' for \textit{christian}).
This shows that \toolname{} is not only comparable to human annotations but sometimes can even surpass them, echoing the findings from some existing work that LLM annotations can have higher quality than human annotations~\cite{He2023AnnoLLMML}.

\begin{figure}[t]
    \centering
    \includegraphics[width=0.75\linewidth]{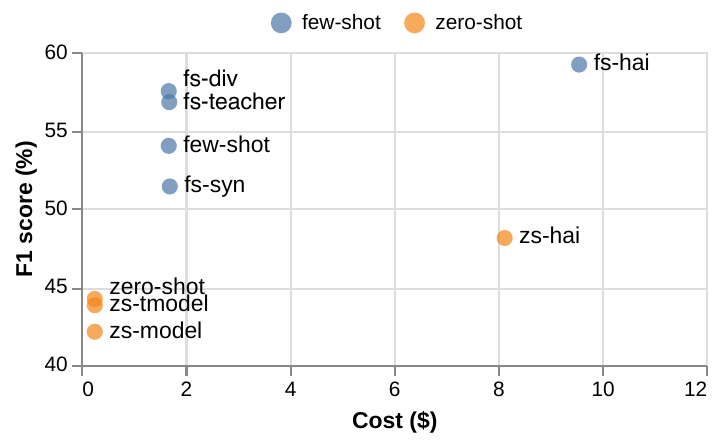}
    \caption{\small We visualize F1 score and cost of each configuration (for CivilComments, n=6k), which shows a clear trend of two trade-offs: whether to use few-shot examples (higher \DGone{} with higher \DGthree{}), and whether to have human in the loop (higher \DGone{} with more \DGtwo{}).}
    \label{fig:acc_cost}
    \vspace{-10pt}
\end{figure}

\subsection{RQ2: Cost and Latency}
\paragraph{Setup}
In our experiments, we use a local machine with 2 A6000 GPUs for running \texttt{flan-t5-xxl} and use OpenAI's APIs to query \texttt{gpt-4-turbo-preview}.
We collected the end-to-end slicing latency and calculated the number of annotations per second as the normalized latency.
To measure cost, we collected the number of input/output tokens for each LLM query and estimated the cost (in USD) based on charges from LLM providers. 
We estimated the human cost based on the average pay for crowdworkers and data scientists.
The calculation details can be found in the appendix.\footnote{\url{https://github.com/malusamayo/SemSlicer/blob/main/appendix.md}}

\paragraph{Result: \toolname{} can slice a dataset with 6,000 rows in 13.5 minutes with a cost of \$1.70 per slice in the most accurate automated configuration.}

We found \toolname{} can annotate 6 to 14 examples per second (Table~\ref{tab:compute}). 
This translates into using 7.1 minutes (zero-shot) to 13.5 minutes (few-shot) for annotating one slice in a dataset of size of 6k (CivilComments).
We estimate that the associated cost is about \$0.26 (zero-shot) to \$1.70 (few-shot) per slice for automated configurations, and up to \$9.56 (M$_\text{fs-hai}$) for human-in-the-loop configurations (Table~\ref{tab:exp_config}).
In contrast, using crowdworkers for the same task would cost \$432.00, which is 44x to 1661x more than \toolname{} but not necessarily more accurate. %

\paragraph{Result: \toolname{} allows flexible trade-offs between design dimensions through different configurations.}
Comparing different configurations, we observed there are clear trade-offs between design dimensions among different configurations (Figure~\ref{fig:acc_cost}).
Using few-shot examples can increase the cost by around \$1.40 per slice and takes 91\% more time (Table~\ref{tab:compute}), with significant accuracy improvement by 7\% on average (\DGone{} vs. \DGthree{}/\DGfour{}).
Human interventions can increase the cost a lot (\$7.87 per slice), but also come with significant accuracy improvement by 4.3\% to 8.9\%  (\DGone{} vs. \DGtwo{}).

\paragraph{Discussion}
{As shown in} Figure~\ref{fig:acc_cost}
{, there are trade-offs to make between different design dimensions.
In production, the best configuration will depend on the dataset and the slicing criterion difficulty. 
For easier criteria, a simple zero-shot configuration would already work pretty well (e.g., the ``price'' slice in} Table~\ref{tab:result} has a 95.9\% F1-score with M$_{\text{zero-shot}}$). 
For more difficult criteria, we recommend users to select configurations based on the following two principles:

\begin{itemize}
    \item {If there is more} \DGthree{} {budget, always try few-shot examples. Use a teacher model to generate example labels if available.}
    \item {If there is more} \DGtwo{} {budget, allocate it to instruction refinement.}
\end{itemize}

\subsection{RQ3: Usefulness}
\label{sec:eval-rq3}
We approach usefulness from two perspectives, by understanding (1) whether \toolname{} can help identify \textit{known} under-performing slices 
and (2) whether \toolname{} can help practitioners conduct model evaluation and generate \textit{additional} insights.

For the first part, we demonstrate usefulness if \toolname{} can reliably detect under-performing slices, a common use of data slicing (cf. Section~\ref{sec:background}).
For the second part, we conduct a human-subject case study to triangulate the automated experiments, by showing that practitioners found \toolname{} useful in their real-world cases.

\subsubsection{Can \toolname{} help identify under-performing
slices?}

\paragraph{Setup}
A common use of slicing is to identify parts of the input space where the model under-performs. That is, we want to identify the slices in which their downstream \textit{task performance} (i.e., the accuracy of the original task, such as toxicity detection) is statistically significantly worse than the average, even if the \textit{slicing accuracy} is not perfect.
Specifically, we want to evaluate whether our slicing accuracy is good enough to detect those same slices as under-performing that we would have detected with perfect slicing accuracy from the ground-truth attributes in our datasets.

To understand whether \toolname{} can identify under-performing slices, we measure \textit{task performance} for each dataset or slice.
For AdaTest, we reuse the reported predictions from their commercial sentiment analysis model and compare them against the labels in the dataset.
For HateCheck, Amazon, CivilComments, we use popular models (listed in the appendix) from Hugging Face and compare their predictions against the labels in the dataset.
For all datasets, we report \textit{task performance} with the standard accuracy metric.

For our analysis, we determine whether a slice (ground truth or computed with \toolname{}) under-performs by comparing the slice's task performance against the overall task performance on the entire dataset, using Fisher's exact test to test whether the difference is statistically significant (p-value $\le0.05$). We then compare under-performing slices detected through \toolname{} with those detected on ground-truth slices. For this experiment, we use the most accurate automated configuration, M$_{\text{fs-teacher}}$.

\paragraph{Result: \toolname{} identifies all 7 under-performing slices.}

We found that 7 out of 26 slices are statistically significantly under-performing and \toolname{} is able to identify all of them (Table~\ref{tab:result}). In addition, \toolname{} identifies only one slice as under-performing that is not under-performing according to the ground-truth slice (slice \textit{``gay''} in HateCheck;  false positive).
This demonstrates that \toolname{}'s slicing accuracy is sufficient for the practical application of identifying under-performing slices.

\subsubsection{Can \toolname{} help practitioners conduct model evaluation?}

\paragraph{Setup}
We approached researchers in our contact network to identify ML-related projects and selected one participant. 
The participant worked on a project to understand how LLM annotations align differently with different demographic groups. 
More specifically, the participant had annotators annotate the Social Acceptability~\cite{forbes-etal-2020-social} dataset and computed the correlations between model and human annotations on various slices of \textit{annotator attributes}.
The participant observed that LLMs (e.g., GPT-4) aligned better with Western, White, college-educated, and younger populations and was interested in using \toolname{} to tie correlations back to specific slices on \textit{input text},
which they could not have done without support from semantic data slicing.

Before the study, we first discussed with the participant about what kinds of input text slices they would be interested to see for their dataset.
The participant mentioned that slices concerning different demographics in the input text would be interesting---following this, we collected 6 slicing criteria for our case study (\textit{``gender'', ``age'', ``nationality'', ``education'', ``religion'', ``ethnicity''}).
We then used \toolname{} to generate the slices, using the most accurate automated setup of M$_{\text{fs-teacher}}$.
Finally, we invited the participant for a one-hour study session, where the participant inspected the generated slices (visualized in Zeno~\cite{zeno}) and their model-human correlation scores in think-aloud mode.
We collected their verbalized thoughts and feedback on the slices generated by \toolname{} in the process.

\paragraph{Result: \toolname{} supports new opportunities in model evaluation and helps generate additional insights}
Looking at the slices, the participant quickly realized there was a gap between what they expected and what the slices showed.
When inspecting the slice on gender, the participant found many examples are only superficially related to gender by mentioning gendered pronouns (which is the expected behavior of \toolname{}) but what they really cared about are the examples showing gender-related power imbalance.
With this realization, the participant suggested that we can instead slice on \textit{``power imbalance''} as slicing criteria, and its gender-related or age-related subsets.
This anecdote reflects on a larger theme of how users often need to iterate on their slicing criteria, after realizations of their hidden requirements.

We reran \toolname{} to produce three slices (took 7 minutes end-to-end) and reached back to the participant.
The participant found that, this time, they can agree with most of the examples in the slices.
In a few cases, \toolname{} even nudged the participant to understand the nuances: \textit{``[now] I could see it, especially the mother-in-law thing.''}
Further, the participant was able to generate \textit{additional} insights that they could not have done in their original analysis.
For example, the participant found that for text related to age-related power imbalance, the model aligns best with millennials, with a stark correlation drop for age groups above 40, which they did not observe in the overall dataset.
Overall, the participant found \toolname{} useful to help them develop a more fine-grained understanding of their original results.

\section{Related Work}
In Section~\ref{sec:background}, we already discussed the most closely related work on data slicing. 
Here, we \textit{additionally} discuss related work on \textit{automated prompt engineering} relevant to how we implemented \toolname{}, as well as how \textit{slice discovery} can augment semantic slicing. %

\subsection{Automated Prompt Engineering}
\paragraph{Instructions}
Automated improvement of prompt instructions is an emerging direction that has attracted lots of attention.
~\citet{zhou2023large} generate and paraphrase instruction candidates using LLMs and select instructions based on scores on labeled evaluation data.
~\citet{pryzant-etal-2023-automatic} improve instructions iteratively with erroneous examples.
Both studies assume access to a labeled dataset.
In contrast, \toolname{}'s instruction generation and refinement are designed for a zero-label setting. 

\paragraph{Few-shot examples}
Few-shot examples are known to have a large impact on prompt performance~\cite{pmlr-v139-zhao21c}, and few-shot example selection has been an active field of research.
~\citet{Liu2021WhatMG} propose to retrieve examples that are semantically \textit{similar} to a test sample.
Other work found the retrieved examples are often redundant and proposed to find \textit{diverse} examples with high coverage~\cite{ye-etal-2023-complementary, gupta-etal-2023-coverage}. 

Most of the existing work assumes access to a labeled training dataset, while \toolname{} requires few-shot example selection without labels. 
Even though we cannot use these methods directly, they still inspired our design of the input sampler, where we provide the option to retrieve diverse inputs.

Closest to our work is universal self-adaptive prompting~\cite{wan-etal-2023-universal}, where they select high-confidence examples (based on the self-predictions) as few-shot examples. 
Our early experimentation did not observe consistent improvement from this method and hence we did not report it in our evaluation. 
However, it is still available as an option for input sampling.

DSPy~\cite{khattab2023dspy} is a framework that can automatically generate and optimize prompts for generic LLM pipelines.
\toolname{} is designed specifically for data slicing as a standalone framework, but can potentially be implemented using existing frameworks like DSPy.

\paragraph{Prompt selection.}
Another line of work on zero-label prompt selection~\cite{yang2024improving} aims to select a good prompt from multiple candidates.
Existing work has explored selection using pseudo labels from prompt ensembles~\cite{liao2022zerolabel}, perplexity scores~\cite{gonen-etal-2023-demystifying}, and mutual information~\cite{sorensen-etal-2022-information}.
Our early experimentation suggested limited improvement from these methods, and hence we did not use them in the final design.

\subsection{Automated Slice Discovery}
Automated slice discovery~\cite{dEon2021TheSA, eyuboglu2022domino} is an error analysis approach based on the idea of automatically identifying under-performing areas in the input space, often using clustering or dimensionality reduction techniques.
These areas can be interpreted as slices and this line of work explores ways to identify meaningful names for those areas,
which could then be considered as slicing criteria.
However, these methods are designed to discover \textit{under-performing} areas, without considering whether these are \textit{complete} according to some human-interpretable abstraction.
As pointed out by \citet{Johnson2023WhereDM}, their produced slices can be misleading, in that models that under-perform on the found slice often do not under-perform on what would be a complete slice for that concept.
In contrast, \toolname{} aims to identify \textit{all} relevant examples for a given slicing criteria.
Overall automated slice discovery is complementary to our approach as an \textit{error analysis} technique to generate hypotheses (cf. Fig.~\ref{fig:eval_flow}), whereas we focus on creating slices to \textit{validate hypotheses}.

\section{Conclusion}
In this work, we present \toolname{}, a framework that supports semantic data slicing. 
We provide a comprehensive view of how data slicing is broadly applicable in ML engineering and demonstrate how \toolname{} provides new opportunities missed by existing programmatic slicing methods.
Our evaluation found \toolname{} can produce accurate slices at a low cost, with flexible trade-offs among different design dimensions, and is useful for model evaluation.

{Insights from our work provide new research opportunities:}

\textit{Semantic slicing at a reduced cost.}
In its current design, \toolname{} {costs \$0.26 to \$1.70 to automatically produce slices from a dataset of 553k tokens.
This is much cheaper than recruiting crowdworkers but can still incur substantial cost for bigger datasets, longer inputs, or more slices.
One promising direction is to explore retrieval-based methods}~\cite{khattab2020colbert} {to filter out irrelevant data points and reduce the annotations needed.
Another is to train small customized models specifically for the task of data slicing}.

\textit{Semantic slicing with higher accuracy.}
\toolname{} {has implemented a wide range of configuration options.
However, the field of automated prompt engineering is evolving quickly.
Future work can incorporate additional emerging research findings and test their combinations.
For example, one direction is to build an agentic framework}~\cite{xi2023rise} {for semantic slicing,
where the slicing agent actively refines instructions and few-show examples, based on automated or LLM-generated feedback.}

\textit{Interactive semantic slicing.}
 The current design of \toolname{} {allows human intervention through editing slicing conditions and post-editing instructions.
Future work can integrate} \toolname{} in an interactive interface (e.g., Zeno~\cite{zeno}), {and explore different interaction designs.
For example, humans can annotate few-shot examples in an active learning loop}~\cite{settles2009active}, {or provide verbalized feedback for instruction refinement.
This way, humans have more agency in prompt construction, potentially leading to faster iterations.}

\begin{acks}
We thank Claire Le Goues, Graham Neubig, Alex Cabrera, Xinran Zhao, Jenny Liang, Vijay Viswanathan, Manisha Mukherjee,  and others for their feedback on this work.
Kaestner and Yang's work was, in part, supported by NSF 2106853, 2131477, and 2206859.
Yining's work was supported by the Top Open Program at Tsinghua University as an REU student at Carnegie Mellon.
Lewis' work was funded and supported by the Department of Defense under Contract No. FA8702-15-D-0002 with Carnegie Mellon University for the operation of the Software Engineering Institute, a federally funded research and development center (DM24-1142).
The work was also supported by gift funds from Amazon, Google Research, and Adobe Research.
\end{acks}

\bibliographystyle{ACM-Reference-Format}
\bibliography{main}

\end{document}